\documentclass[12pt]{article}
\usepackage{graphicx}
\usepackage{amssymb}
\usepackage{amsmath}

\begin{document}

\begin{titlepage}

\begin{flushright}
ICRR-Report-618-2012-7\\
IPMU 12-0148
\end{flushright}

\vskip 1.35cm

\begin{center}

{\large 
{\bf Remarks on Hubble Induced Mass from Fermion Kinetic Term} 
}

\vskip 1.2cm

Masahiro Kawasaki$^{a,b}$
and
Tomohiro Takesako$^a$ \\

\vskip 0.4cm

{ \it$^a$Institute for Cosmic Ray Research,
University of Tokyo, Kashiwa 277-8582, Japan}\\
{\it $^b$Kavli Institute for the Physics and Mathematics of the Universe,
University of Tokyo, Kashiwa 277-8568, Japan}\\

\date{\today}

\begin{abstract} 
We evaluate the effective mass of a scalar field 
which interacts with visible sector via Planck-suppressed coupling in supergravity framework.
We focus on the radiation-dominated (RD) era after inflation and the contribution from a fermionic field in the thermal bath.
We find that, in RD era, the fermion kinetic term gives the effective mass of the order of Hubble scale to the scalar field.
\end{abstract}

\end{center}
\end{titlepage}

\section{Introduction}
\label{sec:intro}
Supersymmetry provides a key for inflation model building since supersymmetric models have many scalar fields~\cite{Lyth:1998xn}.
Its local version, supergravity, also provides many interesting phenomena.
In particular, supergravity effect generally gives the effective mass of the order of Hubble scale $H$ to all scalar fields during inflation~\cite{hep-ph/9405389}.
Thus, even if an originally massless scalar field $\phi$ has Planck-suppressed couplings only, a large effective mass is generated for $\phi$.

Recently, it has been shown that the effective mass of the order of $H$ for such a Planck-suppressed interacting scalar field $\phi$ is also generated during the radiation-dominated (RD) era~\cite{Kawasaki:2011zi}. 
However, in Ref.~\cite{Kawasaki:2011zi}, only a scalar field in the thermal bath is considered as a source for the effective mass of $\phi$.
One would expect a fermion field in the thermal bath also gives the effective mass of the order of $H$ to $\phi$.
In this study, we confirm this observation based on thermal field theory.

The construction of this study is as following:
in Section~\ref{sec:Fermion}, we formulate the contribution from a fermion kinetic term to the effective mass for the Planck-suppressed interacting scalar field $\phi$, 
assuming the fermionic field is in the thermal bath.
Then, we evaluate the effective mass based on thermal field theory.
Section~\ref{sec:conc} is devoted to conclusion.

\section{Contribution from a fermion kinetic term}
\label{sec:Fermion}
In this section, we formulate the effective mass for the Planck-suppressed interacting scalar field $\phi$.
We focus on the contribution from the kinetic term of a fermion in the thermal bath.

\subsection{Formulation of the effective mass}
We consider a scalar field $\phi$ and a chiral fermion $\tilde \chi$ (we use two-component notation) in supergravity framework.
We assume that masses of $\phi$ and $\tilde \chi$ are originally ({\it i.e.}, at zero temperature) much smaller than the Hubble scale $H$,
and that $\phi$ and $\tilde \chi$ are interacting only via the non-minimal K$\ddot{\text{a}}$hler potential given by
\begin{equation}\label{eq:Kahler}
\begin{split}
K = |\Phi|^2 + |\chi|^2 + c~\frac{|\Phi|^2 |\chi|^2}{M_{\text{P}}^2},
\end{split}
\end{equation}
where $M_{\text{P}} \simeq 2.4 \times 10^{18}~\mathrm{GeV}$ is the reduced Planck mass and $c = \mathcal{O} (1)$ is a model-dependent parameter.
Here, $\Phi$ and $\chi$ are superfields which include the scalar field $\phi$ and the chiral fermion $\tilde \chi$, respectively.
Then, the kinetic term of $\tilde \chi$ is given by~\cite{350988}
\begin{equation}\label{eq:kin}
\begin{split}
\mathcal{L}_{\text{kin.}}^{\tilde \chi}
&= \left( 1 + c \frac{|\phi|^2}{M_{\text{P}}^2}\right) \tilde \chi (x) i \sigma^{\mu} \partial_{\mu} \tilde \chi^* (x),
\end{split}
\end{equation}
where $\sigma^{\mu} = (1, \sigma^i)$ ($\sigma^i$ are the Pauli matrices).
In the following, we consider the effective mass-squared for the scalar field $\phi$, $\tilde m_{\phi}^2$, in RD era.
From the kinetic term Eq.~(\ref{eq:kin}), the effective mass-squared $\tilde m_{\phi}^2$
from the $\phi$ - $\tilde \chi$ Planck-suppressed interaction is generally written as
\begin{equation}\label{eq:formula}
\begin{split}
\tilde m_{\phi}^2 |_{\text{kin.}}^{\text{fermion}}
&= - \frac{c}{M_{\text{P}}^2} \langle \tilde \chi (x) i \sigma^{\mu} \partial_{\mu} \tilde \chi^* (x) \rangle.
\end{split}
\end{equation}
In this study, we assume the chiral fermion $\tilde \chi$ is in thermal equilibrium.
Here and hereafter, $\langle \cdots \rangle = \mathrm{tr} (\mathrm{e}^{- \beta H} \cdots) / \mathrm{tr} (\mathrm{e}^{- \beta \hat{\mathcal{H}}})$ represents the thermal expectation value,
where $\beta = 1/T$ is the inverse temperature and $\hat{\mathcal{H}}$ is the Hamiltonian of the system.
In the next subsection, we evaluate Eq.~(\ref{eq:formula}) using thermal field theory.

\subsection{Evaluation of the expectation value under quasi-particle approximation}
Below, we assume space-time homogeneity and spatial isotropy of the background metric.
We also take the chiral fermion zero-temperature mass $m_0 = 0$ for simplicity, although the following argument can be applied for $m_0 \ll m_f$ ($m_f$ is the thermal mass for the fermion $\tilde \chi$)\footnote{
When the zero-temperature mass is relatively large, $m_0 \simeq m_f ( \ll T)$, we have to reconsider the following discussion.
On the other hand, as we assume the chiral fermion is in thermal equilibrium, $m_0 \gtrsim T$ case is irrelevant here. 
}.
Since the thermalization rate of the fermion is much larger than the Hubble expansion rate,
we evaluate the expectation value in Eq.~(\ref{eq:formula}) in Minkowski space-time in the following discussion.
The Hubble expansion rate relates to the evaluation only through the thermal bath temperature $T$.
Moreover, assuming the thermal bath is large enough, we neglect the backreaction of $\phi$-$\tilde \chi$ interaction to the bath.

First of all, we note the following equation:
\begin{equation}\label{eq:fermion-kin}
\begin{split}
\langle \tilde \chi (x) i \sigma^{\mu} \partial_{\mu} \tilde \chi^* (x) \rangle
&= - i \sigma^{\mu}_{\alpha \dot \alpha} \partial_{\mu}^{x_1} \left( \tilde \Delta^{(+) \dot \alpha \alpha} (x_1, x_2) + \frac{i}{2} \tilde \Delta^{(-) \dot \alpha \alpha} (x_1, x_2) \right) {\bigg |}_{x_1 = x_2 = x},
\end{split}
\end{equation}
where $\alpha, \dot \alpha$ are the spinor indices.
Here, we have defined the correlation functions $\tilde \Delta^{(\pm) \dot \alpha \alpha} (x_1, x_2)$ as 
\begin{equation}
\begin{split}
&\tilde \Delta^{(+) \dot \alpha \alpha} (x_1, x_2) = \frac{1}{2} \langle \left[ \tilde \chi^{* \dot \alpha} (x_1), \tilde \chi^{\alpha} (x_2) \right] \rangle,\\
&\tilde \Delta^{(-) \dot \alpha \alpha} (x_1, x_2) = i \langle \left\{ \tilde \chi^{* \dot \alpha} (x_1), \tilde \chi^{\alpha} (x_2) \right\} \rangle.
\end{split}
\end{equation}
Here, $\tilde \Delta^{(-) \dot \alpha \alpha} (x_1, x_2)$ is called spectral function.
Since the chiral fermion $\tilde \chi$ is in thermal equilibrium, the correlation functions depend only on the difference $x_1 - x_2$:
$\tilde \Delta^{(\pm) \dot \alpha \alpha} (x_1, x_2) = \tilde \Delta^{(\pm) \dot \alpha \alpha} (x_1 - x_2)$.
Thus, applying spatial Fourier transform, we obtain the following expression:
\begin{equation}
\begin{split}
&\langle \tilde \chi (x) i \sigma^{\mu} \partial_{\mu} \tilde \chi^* (x) \rangle\\
&= - i \sigma_{\alpha \dot \alpha}^{\mu} \partial_{\mu}^{x_1} \int \frac{\mathrm{d}^3 \bold{p}}{(2 \pi)^3}~\mathrm{e}^{i \bold{p} \cdot (\bold{x_1 - x_2})} 
\left( \tilde \Delta^{(+) \dot \alpha \alpha }_{\bold{p}} (t_1- t_2) + \frac{i}{2} \tilde \Delta^{(-) \dot \alpha \alpha}_{\bold{p}} (t_1- t_2) \right) {\bigg |}_{x_1 = x_2 = x} \\
&= \int \frac{\mathrm{d}^3 \bold{p}}{(2 \pi)^3}~\mathrm{tr} \left\{ \left(- i \partial_y + \bold{p} \cdot \boldsymbol{\sigma} \right)
\left( \tilde \Delta^{(+)}_{\bold{p}} (y) + \frac{i}{2} \tilde \Delta^{(-)}_{\bold{p}} (y) \right) \right\} {\bigg |}_{y=0},
\end{split}
\end{equation}
where $y = t_1 - t_2$.
To go further, we can use the following KMS relation between the equilibrium fermion correlation functions~\cite{8651}:
\begin{equation}
\begin{split}
\tilde \Delta^{(+)}_{\bold{p}} (\omega) = - \frac{i}{2} \tanh \left( \frac{\beta \omega}{2} \right) \tilde \Delta_{\bold{p}}^{(-)} (\omega).
\end{split}
\end{equation}
From this relation, we obtain
\begin{equation}\label{eq:delta+-}
\begin{split}
\tilde \Delta^{(+)}_{\bold{p}} (y) + \frac{i}{2} \tilde \Delta^{(-)}_{\bold{p}} (y)
&= \int_{- \infty}^{\infty} \frac{\mathrm{d} \omega}{4 \pi}~\mathrm{e}^{- i \omega y} \left( \tanh \left( \frac{\beta \omega}{2} \right) - 1 \right) \tilde \rho_{\bold{p}} (\omega),
\end{split}
\end{equation}
where we have used $\tilde \Delta^{(-)}_{\bold{p}} (\omega) = i \tilde \rho_{\bold{p}} (\omega)$.

Now, we are in position to use the formula for the spectral function $\tilde \rho_{\bold{p}} (\omega)$ under quasi-particle approximation.
When we take the zero width limit, the spectral function $\tilde \rho_{\bold{p}} (\omega)$ is given by~\cite{Bellac}
\begin{equation}\label{eq:quasi-spec}
\begin{split}
\tilde \rho_{\bold{p}} (\omega)
&= \pi \left[ Z_+ (p) \delta (\omega - \Omega_+ (p)) +  Z_- (p) \delta (\omega + \Omega_- (p)) \right] \left( 1 + \hat{\bold{p}} \cdot \boldsymbol{\sigma} \right) \\
&~~~ + \pi \left[ Z_- (p) \delta (\omega - \Omega_- (p)) +  Z_+ (p) \delta (\omega + \Omega_+ (p)) \right] \left( 1 - \hat{\bold{p}} \cdot \boldsymbol{\sigma} \right),
\end{split}
\end{equation}
where $Z_{\pm} ( p )$ is the residue of the pole $\Omega_{\pm} ( p )$ and $\hat{\bold{p}} = \bold{p} / |\bold{p}|$.
The brief derivation of Eq.~(\ref{eq:quasi-spec}) and the limiting formulae for $Z_{\pm} ( p ), \Omega_{\pm} ( p )$ are summarized in Appendix.
Using Eqs.~(\ref{eq:delta+-}), (\ref{eq:quasi-spec}), Eq.~(\ref{eq:fermion-kin}) is reduced to a relatively simple form as following:
\begin{equation}\label{eq:kin-exact}
\begin{split}
\langle \tilde \chi (x) i \sigma^{\mu} \partial_{\mu} \tilde \chi^* (x) \rangle
&= \frac{1}{\pi^2} \int_0^{\infty} \mathrm{d} p~p^2 {\Bigg \{} Z_+ (p) \left( \Omega_+ (p) - p \right) \left( n_F (\Omega_+) - \frac{1}{2} \right) \\
&~~~~~~~~~~~~~~~~~~~~~~~~~+ Z_- (p) \left( \Omega_- (p) + p \right) \left( n_F (\Omega_-) - \frac{1}{2} \right)  {\Bigg \}},
\end{split}
\end{equation}
where $n_F (\omega) = 1 / (\mathrm{e}^{\beta \omega} + 1)$ is the Fermi-Dirac distribution function.

To proceed the analysis further, let us approximate $n_F (\Omega_{\pm} (p)) = 1 / (\mathrm{e}^{\beta \Omega_{\pm} (p)} + 1)$.
The approximation formula we use here is based on the following three observations:
first, $n_F (\Omega_{\pm} (p))$ has a cutoff around $\Omega_{\pm} \simeq T$.
Second, when $\Omega_{\pm}$ is small (when $p \ll T$), we can neglect $\Omega_{\pm} (p)$ dependence of $n_F (\Omega_{\pm} (p))$.
Finally, $\Omega_{\pm} (p) \simeq p$ for $p \gg m_f$, where $m_f^2  = \kappa' g^2 T^2$ is the thermal mass for the chiral fermion $\tilde \chi$.
Here, $g$ is a coupling constant and $\kappa' \lesssim \mathcal{O} (1)$ is a model-dependent constant\footnote{
For example, if the interaction term is given by $\mathcal{L}_{\text{int.}} = - g \varphi \tilde \chi \tilde \lambda + h.c.$ ($\varphi$ is a complex scalar field and $\tilde \lambda$ is a chiral fermion),
we obtain $\kappa' = 1/16$ under the one-loop Hard Thermal Loop approximation~\cite{Bellac}.
}.
From the above observations, it is reasonable to use the following approximation formula for $n_F (\Omega_{\pm} (p))$ for all intervals of $p$:
\begin{equation}
\begin{split}
n_F (\Omega_{\pm} (p)) \simeq \frac{1}{\mathrm{e}^{\beta p} + 1}.
\end{split}
\end{equation}
Then, the expectation value for the kinetic term Eq.~(\ref{eq:kin-exact}) becomes
\begin{equation}\label{eq:integ}
\begin{split}
\langle \tilde \chi (x) i \sigma^{\mu} \partial_{\mu} \tilde \chi^* (x) \rangle
&= \frac{1}{\pi^2} \int_0^{\infty} \mathrm{d} p~p^2~\left\{ Z_+ (p) \left( \Omega_+ (p) - p \right) + Z_- (p) \left( \Omega_- (p) + p \right) \right\} \frac{1}{\mathrm{e}^{\beta p} + 1},
\end{split}
\end{equation}
where we have neglected the "vacuum" contribution, which is independent of the distribution function, for simplicity\footnote{
The "vacuum" contribution may lead to $\langle \tilde \chi (x) i \sigma^{\mu} \partial_{\mu} \tilde \chi^* (x) \rangle_{\text{vac}}
\simeq \frac{-1}{2 \pi^2} \int_{0}^{M_{\text{P}}} \mathrm{d} p~p^2 \times \frac{m_f^2}{p} = - \frac{m_f^2 M_{\text{P}}^2}{4 \pi^2}$,
where we have introduced the cut-off scale $M_{\text{P}}$.
Unfortunately, this is $T$-dependent quadratic divergence, though we do not expect the chiral fermion kinetic term has such a huge expectation value.
Thus, for simplicity, we neglect the "vacuum" contribution in this study.
}.
The contributions from the integration intervals $[m_f, T]$ and $[0, m_f]$ in Eq.~(\ref{eq:integ}) give the leading and the next leading order contributions  in terms of the coupling $g$, respectively.
Thus, assuming the coupling $g$ is small enough, we apply the approximation formulae $Z_{+} \simeq 1,~Z_- \simeq 0,~\Omega_{+} (p) \simeq p + m_f^2 /p,~\Omega_{-} (p) \simeq p$
(see Eq.~(\ref{eq:limit})) to the whole interval $[0, \infty]$ although these formulae are valid only within the interval $[m_f, T]$:
\begin{equation}\label{eq:res-kin-fermi}
\begin{split}
\langle \tilde \chi (x) i \sigma^{\mu} \partial_{\mu} \tilde \chi^* (x) \rangle
&\simeq \frac{1}{\pi^2} \int_{0}^{\infty} \mathrm{d} p~p^2 \times \frac{m_f^2}{p} \frac{1}{\mathrm{e}^{\beta p} + 1} = \frac{m_f^2 T^2}{12}.
\end{split}
\end{equation}
From Eq.~(\ref{eq:res-kin-fermi}), the effective mass of the scalar field $\phi$ originated from the kinetic term of the chiral fermion  is given by
\begin{equation}\label{eq:main-res}
\begin{split}
\tilde m_{\phi}^2 |_{\text{kin.}}^{\text{fermion}}
&\simeq - \frac{c}{M_{\text{P}}^2} \frac{m_f^2 T^2}{12}\\
&= - \frac{15 c \kappa'}{2 \pi^2 g_*} g^2 H^2,
\end{split}
\end{equation}
where $m_f^2 = \kappa' g^2 T^2$ and the relation $3 M_{\text{P}}^2 H^2 = \frac{\pi^2 g_*}{30} T^4$ in RD era is used,
and $g_*$ is the relativistic degrees of freedom in the thermal bath.
Eq.~(\ref{eq:main-res}) is our main result in this study.

\section{Conclusion}
\label{sec:conc}
We have evaluated the effective mass of a scalar field $\phi$ which interacts only via Planck-suppressed operator to visible sector given in Eq.~(\ref{eq:Kahler}).
We focus on RD era and the contribution from a chiral fermion $\tilde \chi$ in the thermal bath.
To make the analysis reliable, we base on thermal field theory for the chiral fermion $\tilde \chi$.
We find that the chiral fermion kinetic term gives the effective mass of the order of $H$ to the scalar field $\phi$ under the quasi-particle approximation for $\tilde \chi$.
The main result in this study is given in Eq.~(\ref{eq:main-res}).
This Hubble-induced mass has almost the same magnitude as the one from scalar field kinetic term in the thermal bath~\cite{Kawasaki:2011zi}.
Such an effective mass of the order of $H$ for the Planck-suppressed interacting scalar field in RD era may affect some cosmological scenarios.
For example, the Hubble-induced mass in RD era would play an important role to solve the cosmological moduli problem~\cite{Linde:1996cx}.
On the other hand, when we consider the curvaton scenario~\cite{Lyth:2001nq} in the framework of supergravity,
the Hubble-induced mass in RD era as well as during inflation will be the main obstacle to the model building.
From our study, it is confirmed that the fermions in the thermal bath contribute to the Hubble-induced mass as well as the scalars in the bath.

\section*{Acknowledgments}
T.T. is grateful to Toru Kojo for a helpful discussion during the conference ``Strong ElectroWeak Matter (SEWM) 2012". 
The work of T.T. is supported in part by JSPS Research Fellowships for Young Scientists.
This work is supported by Grant-in-Aid for
Scientific research from the Ministry of Education, Science, Sports, and
Culture (MEXT), Japan, No.14102004 and No.21111006 (M.~K.)  and also by
World Premier International Research Center Initiative (WPI Initiative), MEXT, Japan.
\section*{Appendix: Spectral function for chiral fermion}
In this Appendix, we briefly review the derivation of the spectral function for the chiral fermion $\tilde \chi$~\cite{Riotto:1996cp, Bellac}.
Let us start with the following thermally corrected self-energy of the chiral fermion:
\begin{equation}
\begin{split}
\Sigma = a - b \hat{\bold{p}} \cdot \boldsymbol{\sigma},
\end{split}
\end{equation}
where $\hat{\bold{p}} = \bold{p}/|\bold{p}|$ ($p_{\mu} = (p_0, \bold{p})$ is the fermion external momentum) and we have neglected the chiral fermion zero-temperature mass.
For the one-loop Hard Thermal Loop approximation, the parameters $a$ and $b$ generally have the following form~\cite{Riotto:1996cp, Bellac}:
\begin{equation}
\begin{split}
&a = \frac{m_f^2}{2 p} \ln \left( \frac{p_0 + p}{p_0 - p} \right), \\
&b = - \frac{m_f^2}{p} \left( 1 - \frac{p_0}{2 p} \ln \left( \frac{p_0 + p}{p_0 - p} \right) \right),
\end{split}
\end{equation}
where $m_f$ is the fermion thermal mass.
For instance, if we assume an interaction term $\mathcal{L}_{\text{int}} = - g \varphi \tilde \chi \tilde \lambda + h.c.$ ($\varphi$ is a complex scalar field and $\tilde \lambda$ is a chiral fermion),
we obtain $m_f^2 = g^2 T^2/16$.

The inverse propagator for $\tilde \chi$ including the thermally corrected self-energy is given by
\begin{equation}
\begin{split}
i S^{-1} (P)
&= \sigma^{\mu} p_{\mu} - \Sigma \\
&=  (p_0 - p - a + b) \frac{1 + \hat{\bold{p}} \cdot \boldsymbol{\sigma}}{2}
+ (p_0 + p - a - b) \frac{1 - \hat{\bold{p}} \cdot \boldsymbol{\sigma}}{2}.
\end{split}
\end{equation}
Then, the propagator for $\tilde \chi$ can be written as following:
\begin{equation}
\begin{split}
- i S (P)
&= (p_0 - p - a + b)^{-1} \frac{1 + \hat{\bold{p}} \cdot \boldsymbol{\sigma}}{2}
+ (p_0 + p - a - b)^{-1} \frac{1 - \hat{\bold{p}} \cdot \boldsymbol{\sigma}}{2}.
\end{split}
\end{equation}
Thus, the spectral function for the chiral fermion $\tilde \chi$ is given by
\begin{equation}
\begin{split}
\tilde \rho_{\bold{p}} (\omega)
&= S (\omega + i \epsilon, \bold{p}) - S (\omega - i \epsilon, \bold{p}) \\
&= \rho^{+}_{\bold{p}} (\omega) \frac{1 + \hat{\bold{p}} \cdot \boldsymbol{\sigma}}{2} 
+ \rho^{-}_{\bold{p}} (\omega) \frac{1 - \hat{\bold{p}} \cdot \boldsymbol{\sigma}}{2},
\end{split}
\end{equation}
where
\begin{equation}
\begin{split}
\rho^{\pm}_{\bold{p}} (\omega) = - 2~\mathrm{Im} \frac{1}{(p_0 - a) \mp (p - b)} {\Big |}_{p_0 = \omega + i \epsilon}.
\end{split}
\end{equation}
Here, $\epsilon \to + 0$ is the infinitesimal parameter.

In thermal field theory, the spectral function is known to be divided into quasi-particle pole contribution and continuous state one~\cite{Bellac}:
\begin{equation}
\begin{split}
&\rho^{\pm}_{\bold{p}} (\omega) = 2 \pi \left( \rho^{\pm (\text{pole})}_{\bold{p}} (\omega) + \rho^{\pm (\text{cont.})}_{\bold{p}} (\omega) \right), \\
&\rho^{\pm (\text{pole})}_{\bold{p}} (\omega) = Z_{\pm} (p) \delta (\omega - \Omega_{\pm} (p)) + Z_{\mp} (p) \delta (\omega + \Omega_{\mp} (p)),
\end{split}
\end{equation}
where the pole width is taken to be zero.
Thus, under the quasi-particle approximation, the spectral function is given by
\begin{equation}
\begin{split}
\tilde \rho_{\bold{p}} (\omega)
&= \rho^+_{\bold{p}} (\omega) \frac{1 + \hat{\bold{p}} \cdot \boldsymbol{\sigma} }{2} + \rho^-_{\bold{p}} (\omega) \frac{1 - \hat{\bold{p}} \cdot \boldsymbol{\sigma} }{2} \\
&\simeq \pi  \left(  \rho^{+ (\text{pole})}_{\bold{p}} (\omega) \left(  1 + \hat{\bold{p}} \cdot \boldsymbol{\sigma} \right) 
+ \rho^{- (\text{pole})}_{\bold{p}} (\omega) \left( 1 - \hat{\bold{p}} \cdot \boldsymbol{\sigma} \right) \right) \\
&= \pi \left[ Z_+ (p) \delta (\omega - \Omega_+ (p)) +  Z_- (p) \delta (\omega + \Omega_- (p)) \right] \left( 1 + \hat{\bold{p}} \cdot \boldsymbol{\sigma} \right) \\
&~~~ + \pi \left[ Z_- (p) \delta (\omega - \Omega_- (p)) +  Z_+ (p) \delta (\omega + \Omega_+ (p)) \right] \left(1 - \hat{\bold{p}} \cdot \boldsymbol{\sigma} \right).
\end{split}
\end{equation}
This is what we would like to derive here.
For the sake of convenience, we write down the limiting formulae for the residues $Z_{\pm} (p)$ and the dispersion relations for the poles $~\Omega_{\pm} (p)$
under the one-loop Hard Thermal Loop approximation:
\begin{equation}\label{eq:limit}
\begin{split}
&Z_+ (p) 
= \frac{\Omega_+ (p)^2 - p^2}{2 m_f^2} 
\simeq \begin{cases}
&\frac{1}{2} + \frac{p}{3 m_f} ~~~~~~~~~~~~~~~~~~~~~~ (p \ll m_f) \\
&1 + \frac{m_f^2}{2 p^2} \left( 1 - \mathrm{ln} \frac{2 p^2}{m_f^2} \right) \sim 1 ~~(p \gg m_f)
\end{cases} \\
&Z_- (p) 
= \frac{\Omega_- (p)^2 - p^2}{2 m_f^2} 
\simeq \begin{cases}
&\frac{1}{2} - \frac{p}{3 m_f} ~~~~~~~~~~~~~~~~~~~~~~ (p \ll m_f) \\
& \frac{2 p^2}{\mathrm{e} m_f^2}~\mathrm{exp} \left( - \frac{2 p^2}{m_f^2} \right) \sim 0 ~~~~~~(p \gg m_f)
\end{cases} \\
&\Omega_+ (p) 
\simeq \begin{cases}
&m_f + \frac{1}{3} p ~~~~ (p \ll m_f) \\
&p + \frac{m_f^2}{p} ~~~~~~(p \gg m_f)
\end{cases} \\
&\Omega_- (p) 
\simeq \begin{cases}
&m_f - \frac{1}{3} p ~~~~~~~~~~~~~~~~~~~~ (p \ll m_f) \\
&p + \frac{2 p}{\mathrm{e}}~\mathrm{exp} \left( - \frac{2 p^2}{m_f^2} \right) ~~~~~~(p \gg m_f).
\end{cases}
\end{split}
\end{equation}
We note that, in the originally ({\it i.e.}, at zero-temperature) massless limit, Dirac and Majorana fields have the same dispersion relations as in Eq.~(\ref{eq:limit})~\cite{Riotto:1996cp, Bellac}.

{}

\end{document}